# Robust Automatic Whole Brain Extraction on Magnetic Resonance Imaging of Brain Tumor Patients using Dense-Vnet

Sara Ranjbar[1*], Kyle W. Singleton[1], Lee Curtin[1], Cassandra R. Rickertsen[1], Lisa E. Paulson[1], Leland S. Hu[1,2], J. Ross Mitchell[3§], Kristin R. Swanson[1§]

[1]Mathematical NeuroOncology Lab, Precision Neurotherapeutics Innovation Program, Department of Neurological Surgery, Mayo Clinic, Phoenix, AZ, USA
[2]Department of Diagnostic Imaging and Interventional Radiology, Mayo Clinic, Phoenix, AZ,USA
[3]Department of Biostatistics and Bioinformatics, Moffitt Cancer Center and Research Institute, Tampa, Florida, USA

*Corresponding Author: ranjbar.sara@mayo.edu
§Authors contributed equally.

## Abstract

Whole brain extraction, also known as skull stripping, is a process in neuroimaging in which non-brain tissue such as skull, eyeballs, skin, etc. are removed from neuroimages. Skull striping is a preliminary step in presurgical planning, cortical reconstruction, and automatic tumor segmentation. Despite a plethora of skull stripping approaches in the literature, few are sufficiently accurate for processing pathology-presenting MRIs, especially MRIs with brain tumors. In this work we propose a deep learning approach for skull striping common MRI sequences in oncology such as T1-weighted with gadolinium contrast (T1Gd) and T2-weighted fluid attenuated inversion recovery (FLAIR) in patients with brain tumors. We automatically created gray matter, white matter, and CSF probability masks using SPM12 software and merged the masks into one for a final whole-brain mask for model training. Dice agreement, sensitivity, and specificity of the model (referred herein as DeepBrain) was tested against manual brain masks. To assess data efficiency, we retrained our models using progressively fewer training data examples and calculated average dice scores on the test set for the models trained in each round. Further, we tested our model against MRI of healthy brains from the LBP40A dataset. Overall, DeepBrain yielded an average dice score of 94.5%, sensitivity of 96.4%, and specificity of 98.5% on brain tumor data. For healthy brains, model performance improved to a dice score of 96.2%, sensitivity of 96.6% and specificity of 99.2%. The data efficiency experiment showed that, for this specific task, comparable levels of accuracy could have been achieved with as few as 50 training samples. In conclusion, this study demonstrated that a deep learning model trained on minimally processed automatically-generated labels can generate more accurate brain masks on MRI of brain tumor patients within seconds.

## Acknowledgement

This publication would not have been possible without the support of the James S. McDonnell Foundation, the Ivy Foundation, the Mayo Clinic, the Zicarelli Foundation and the NIH (R01 NS060752, R01 CA164371, U54 CA210180, U54 CA143970, U54 CA193489, U01 CA220378).

## Introduction

Magnetic resonance imaging (MRI) has a pivotal role in non-invasive diagnosis and monitoring of many neurological diseases such as Alzheimer's disease and dementia[1], brain aneurysm, stroke[2], and primary and metastatic brain tumors[3]. The large amount of data produced in routine patient care has prompted the birth of many studies aiming to automate image analysis tasks relevant to patient care including surgical planning, volumetric analyses[4], study of anatomical structures[5], tissue classification[6–10], disease staging[11,12], and localization of pathology[3]. To be successful in characterization of both normal baseline and pathological deviation[13], non-brain tissue (fat, skull, eyeballs, eyes, teeth, etc.) needs to be removed from anatomical MRI. As manual annotation of brain tissue on every slice in a 3D volumetric MRI is excruciatingly labor intensive, many automatic 'whole brain extraction' or 'skull stripping' techniques have been introduced in the literature to tackle this need.

Over the years, many approaches to automatic whole brain extraction have been proposed in the literature. Edge-based skull stripping approaches such as BSE[4] and BEA[14] use predetermined sets of parameters to separate brain and non-brain tissue through the use of morphological or region-growing operations. Intensity-based methods such as SPM2[15] and WAT[16] rely on intensity variations to find the edge of the brain. BET[17,18], BET2[17,18], MLS[19], SMHASS[20] are examples of deformable surface-based methods that use image gradient to fit an active contour/curve to the brain. Atlas-based methods such as MAPS[21] and ANTs[22] define the boundaries of the brain by registering images to one or many atlases for improved accuracy. Patch-based methods such as BEaST[23] and SPECTRE[24] are an extension of atlas-based methods in which image-to-atlas registration is performed on non-local image patches. Hybrid methods integrate several of the above approaches to achieve enhanced results. Examples are HWA[25], McStrip[26], ROBEX[27]. The accuracy and robustness of skull stripping methods are key in their adoption, these two measures often being counter-balanced. Several comparative studies have found hybrid methods to be superior in accuracy at the cost of time-efficiency[28,29]. Intensity-based and edge-based methods tend to be fast because of their simplicity, but their accuracies tend to fluctuate across heterogeneous datasets with varying levels of image resolutions, noise, and artifacts[13]. Atlas-based methods are designed for healthy subjects and thus fail in the presence of large pathological tissue on the image such as diffusely invasive glioblastoma (GBM) tumors. Moreover, GBMs are often localized close to the border of the brain and thus can throw off most skull stripping approaches. Among existing methods, OptiBET –a modified version of BET– has shown robustness with brain pathology[30]. In addition, MONSTR[31,32], a patch-based multi-atlas skull stripping method, demonstrated robustness with images of schizophrenia, traumatic brain injury, and brain tumors.

Recent success of deep learning methods in the ImageNet[33] challenge has made a lasting impact in computer vision and by extension, in biomedical image analysis. Deep convolutional neural networks have shown success in a number of neuroimaging applications such as MR sequence classification[34], prediction of genetic mutation using MRI[35,36], and tumor segmentation[37–39]. Naturally, several works have explored the utility of deep learning approaches in MRI skull stripping, including the works of Salehi et al[40] and Kleesiek at al[41] that have reported high performance on publicly available datasets of normal brains. The input-agnostic fully convolutional network in the works of Kleesiek at al[41] outperformed BET, BEaST, BSE, ROBEX, and HWA.

Few have fully explored the performance of deep learning approaches on brain tumor data. Given the level of variability that we routinely observe in oncology data, namely in terms of image quality and the varied presentation of brain tumors MRI, we adopted a learning-based approach to tackle this task. The contributions of this work are as follows: 1) assessing the performance of Dense-Vnet architecture in MRI skull stripping of brain tumor data, 2) comparison of performance across the Dense-Vnet MRI input type, 3) conducting a data efficiency experiment to assess the effect of train set size on model performance, and finally 4) assessing the performance of a model trained on brain tumor data on a publicly available dataset of healthy subjects.

## Imaging Data

<u>Brain Tumor Data for Training and Testing.</u> The data source in this work was our in-house IRB-approved repository (described in our previous work[34]) which contains over 70,000 serial structural MR studies of 2,500+ unique brain tumor patients acquired across 20+ institutions. The data pertaining to this study included paired pretreatment T1-weighted post injection of gadolinium contrast (T1Gd) and T2-weighted fluid-attenuated inversion recovery (FLAIR) series of 721 adult brain tumor patients. These series were randomly assigned to 586 training, 52 validation, and 96 test sets. Selection of T1Gd and FLAIR sequences was a practical decision due to their higher prevalence in our repository. We also restricted inclusion criteria to only pre-treatment images as treatment often significantly alters brain appearance on MRI. No additional restriction was imposed on data selection criteria. The imaging data was acquired from 1990 to 2016. Due to the retrospective nature of this dataset, the quality and resolution of images varied across the year and institution of image acquisition. Thus, we employed a number of preprocessing steps to harmonize the data including noise reduction with nonlinear curvature-flow noise reduction[42], radiofrequency non-uniformity correction reduced using the N4 algorithm[43], and resizing to a common matrix of 240x240x64 voxels. The SimpleElastix framework[44] was used to rigidly co-register the FLAIR volume to the T1Gd volume within each study. Our imaging repository contains patient information and therefore is subject to HIPAA regulations. Due to the proprietary nature of patient data and patient information that is visible in input images of the network (pre skull stripping), we are not at liberty to freely share data with readers. However, data may be available for sharing upon the request of qualified parties as long as patient privacy and intellectual property interests of our institution are not compromised. Typically, data access will occur through a collaboration and may require interested parties to obtain an affiliate appointment with our institution.

<u>Healthy Patient Data for Testing.</u> The LONI Probabilistic Brain Atlas Project (LBPA40)[29] consists of 40 T1-weighted MRI scans of healthy subjects (20 males, 20 females) and their corresponding manually labeled brain masks. This dataset was used only for model evaluation and was not used during model training.

## Brain Masks
Several whole brain segmentation approaches were implemented to create brain masks for model training, model testing, and for comparison with previously successful skull stripping methods.

<u>SPM12-p Masks for Model Training and Validation.</u> Given the large size of our cohort, it was impractical and time-consuming to manually delineate brains on this dataset. Consequently, we used an automatic method to create brain masks for training our network. We relied on the work of Malone et al.[45] for choosing the appropriate method that can serve as a substitute for manual delineation. Malone et al.[45] compared the performance of several methods for total intracranial volume segmentation on T1-weighted MRI of a 288 patients with Alzheimer's disease using manual labels and suggested the total intracranial volume of SPM12 to be an acceptable substitute for labour-intensive brain masks in multi-centric datasets, even in the presence of neurodegenerative pathology. Statistical Parameter Mapping or SPM[46] is an image analysis software developed at University College London that contains tools for processing positron-emitted tomography (PET), voxel-based morphometry (VBM), electroencephalography (EEG), functional-MRI and MRI data. Given an input MRI, the segmentation procedure in SPM12 (the most recent version of SPM) outputs probability density maps of specific structures within the brain including white matter, gray matter and cerebrospinal fluid (CSF). We used this component within SPM12 to automatically segment gray matter, white matter, and CSF maps from our T1Gd MRIs. These three resulting maps were combined into a single brain probably map and thresholded at 0.7 (empirically decided) to generate a brain mask. Since the presence of tumors (e.g., tumor necrosis) resulted in occasional missing regions inside the combined mask, we performed minimal morphological operations to fill in the holes in the combined brain mask. The final post-processed result (referred to as SPM12-p) was stored as a binary mask and used as labels for training and validation. SPM12 was run in Matlab version 2018a and postprocessing steps resulting in SPM12-p mask were executed in Python version 3.6.6 and SciPy library version 1.0.0. Figure 1 shows an example of this process.

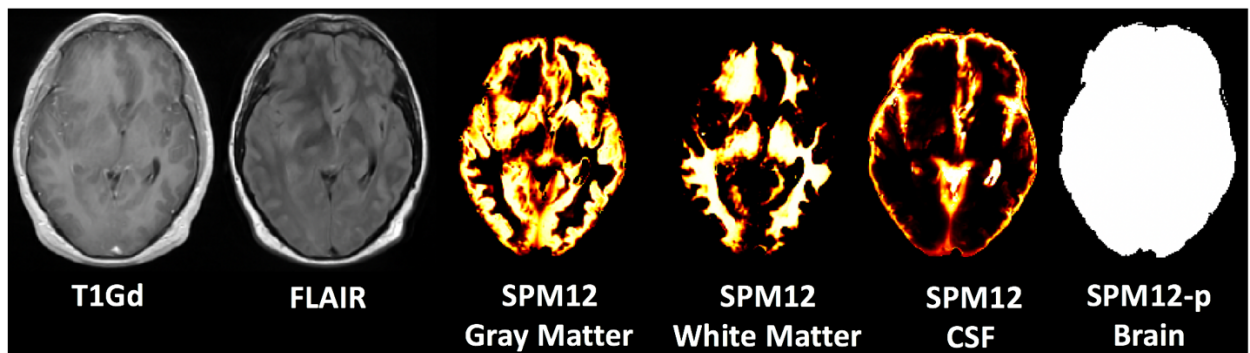

Figure 1 - Gradual steps for creating the brain masks. Images reflect the MRI of a 29 year old male brain tumor patient with a diagnosis of glioblastoma. FLAIR refers to Fluid Attenuated Inversion Recovery (FLAIR) MRI and T1Gd refers to T1-weighted MRI with gadolinium contrast enhancement. Gray matter, white matter, and CSF probability masks were generated using the SPM12 software. Yellow voxels in these masks reflect higher probability. The final brain mask (SPM12-p) was generated by combining SPM12 masks, using a threshold of 0.7, and minimal post-processing.

Manual Masks for Model Testing. To measure model accuracy, we performed manual brain tissue segmentation on 30 randomly-selected test cases. We defined intracranial volume as the combination of gray matter, white matter, subarachnoid CSF, ventricles (lateral, 3rd, 4th), and cerebellum as suggested in the work of Roy et al[31]. Manual segmentation was initiated by two trained individuals with experience in segmentation of tumors on MR imaging data. The segmentation process was initiated with our in-house semi-automatic software used for glioma segmentation. The results were further loaded into the ITK-SNAP[47] software version 3.8.0 (www.itksnap.org) and visually inspected for imperfections and were corrected as needed. This process took about an hour per case. Figure 2 compares the manual brain mask with the automated SPM12-p brain mask for one patient in our test set cohort.

Multi-cONtrast brain STRipping (MONSTR) Masks for Comparison. In addition to the above brain masks, for the 30 test cases processed manually, we also used Multi-cONtrast brain STRipping method (MONSTR)[31] to compare against other methods in the literature. MONSTR[31,32] is a patch-based multi-atlas skull stripping method that has previously demonstrated robustness with MRI of brain tumor patients. MONSTR brain masks were generated using a containerized version of the MONSTR method called from Python 3.6.6 using T1Gd and FLAIR contrasts as inputs.

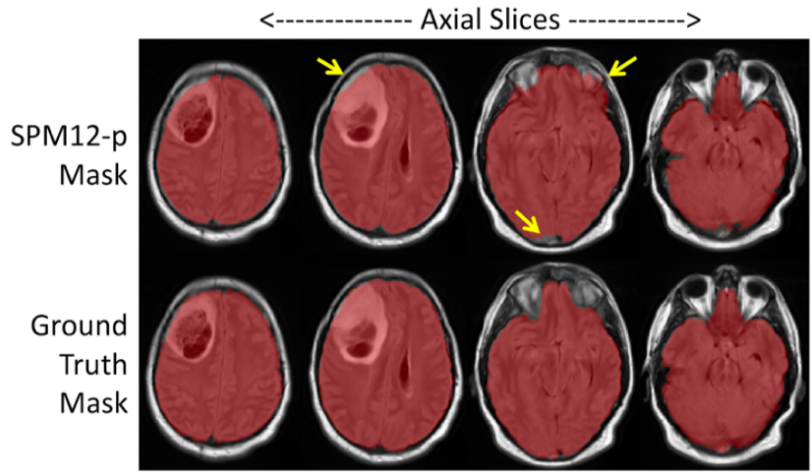

Figure 2 – An example of SPM12-p mask compared to ground truth generated manually. Despite the occasional under- and over-segmentation (arrows), automatically generated brain masks correctly identified brain boundaries even in the presence of a tumor in the brain.

**Network Architecture and Training Approach**

Network Architecture. Training was conducted using the Tensorflow[48]-based deep learning platform, NiftyNet[49,50] version 0.5.0. NiftyNet is a modularly-structured deep learning platform tailored towards medical image analysis applications with modules for pre-processing, network training, evaluation, and inference. For this semantic segmentation task, we used the dense V-network (Dense-Vnet)[51] architecture, a fully connected convolutional neural network[52] that previously has demonstrated success in establishing voxel-voxel connections between input and output images[51]. Dense-Vnet consists of three layers of dense feature stacks[53] whose outputs are concatenated after a single convolution in the skip connection and bilinear upsampling. Supplementary Table 1 presents the setting of parameters in the configuration file used for training a network using the NiftyNet platform. Hereon, we refer to our deep learning model as 'DeepBrain'.

The Main Experiments. Training was conducted over 500 epochs using 586 training and 52 validation samples. No augmentation was performed on our data. During training, model checkpoints were locally saved every 20 epochs. Optimization was implemented using dice loss and adaptive moment estimation (Adam) optimizer[54]. We repeated model training 3 times: using only T1Gd series, only FLAIR series, and both series as inputs. Details of training procedure, network architecture, and parameters were identical between runs. All experiments were conducted in Tensorflow 1.12.0 using an Ubuntu 17.10 system with a single Nvidia TITAN V GPU.

Data Efficiency Experiments. To contribute towards green and reproducible AI[55], we conducted a data efficiency experiment in which we estimated the effect of training set size on model performance. We repeated model training with progressively fewer training samples (500, 400, 300, 200, 100, and 50). Details of training procedure, network architecture, and parameters were identical to the training experiments on the entire cohort. The final model in all experiments was identified among checkpoints by calculating dice loss on the validation set and selecting the model with the best performance.

**Performance Evaluation**

We used all test cases (N=96) to compare train time per iterations, inference time per case, and average dice agreement between predictions and labels (SPM12-p masks). To evaluate performance against ground truth, we used dice overlap coefficient, sensitivity, and specificity to compare the predicted brain masks with manual masks (N=30 out of 96 test cases). Let G be the ground truth image and S the segmentation result. The dice coefficient (D), Sensitivity, and Specificity were defined as follows:

$$Dice\ (G,S) = \frac{2\ TP}{2TP + FP + FN}\ ,\ \ Sensitivity\ (G,S) = \frac{TP}{TP + FN}\ ,\ \ Specificity\ (G,S) = \frac{TN}{TN + FP}$$

where TP, FP, FN are the number of true positive, false positive, and false negative, respectively. Sensitivity measures the detection rate of brain tissue while specificity measures how much non-brain tissue is correctly identified. Finally, the dice score evaluates the trade-off between sensitivity and specificity. Paired t-test was used to compare the results across runs. These performance measures are reported for brain tumor MRIs with available manual masks. Mean and standard deviation of performance measures were calculated to reflect the range of performance. A p-value lower than 0.05 was used to assess statistically significant differences in performance between experiments. All statistical comparisons were performed in Python 3.6.6 using the SciPy[56] package version 1.0.0.

To compare our work with other non-DL skull stripping methods in the literature, we also calculated dice, sensitivity, and specificity of the MONSTR[31] algorithm on our data. For the data efficiency experiment, we report dice scores with training conducted on progressively smaller subsets of the training cohort. Finally, we compared the robustness of our model to other deep-learning skull stripping methods in the literature [41,57,58] using LBPA40 dataset of healthy subjects. Here, we compared dice score, sensitivity, and specificity of our results with others. Reported results for the deep learning methods devised by others were acquired from the respective publications.

## Results
Performance on Brain Tumor Data. The three versions of DeepBrain (trained on T1Gd, FLAIR, and both) yielded similar levels of agreements between predictions and labels. Table 1 compares the performance of DeepBrain across input types on previously unseen test cases. On average, when DeepBrain was trained on FLAIR, it achieved the highest dice and sensitivity while the model trained on both sequences was superior to single input models in specificity (98.84%).

Performance compared to MONSTR and SPM12 on Brain Tumor Data. Table 2 compares the performance of our model with other non-DL brain masks created in this work. While MONSTR did not fail to include the regions occupied by tumors into the segmentation, its performance was much worse in identifying the boundaries of the brain in other regions. We observed over- and under- segmentations in MONSTR-generated brain masks especially at the top and bottom of the brain. In comparison, SPM12-p showed a much-improved sensitivity with statistical significance over MONSTR. DeepBrain was superior in dice score and showed significantly higher sensitivity than both non-DL methods. Figure 3 shows an example of predicted brain masks using DeepBrain for a test case, where our approach performed much better than the other methods. With respect to runtime, we created SPM12-p masks for our cohort within an average of 2-3 minutes which was lower than MONSTR runtime of 10-20 minutes. Longer run time for MONSTR is expected as atlas-based methods tend to take longer than other approaches. The average runtime for DeepBrain was outstandingly faster than the other methods for a mere 2 seconds per case.

Table 1 – Mean and standard deviation of performance for DeepBrain using 30 brain tumor cases with available manual brain mask. Values represent mean and standard deviation of scores.

| CNN Input | Dice score $\mu(\sigma)$ | Sensitivity $\mu(\sigma)$ | Specificity $\mu(\sigma)$ |
|---|---|---|---|
| T1Gd | [ab] 93.09 (1.78) | 96.14 (3.81) | [d] 97.92 (1.28) |
| FLAIR | [a] **94.54** (1.09) | [c] **96.39** (2.34) | 98.48 (1.05) |
| T1Gd+FLAIR | [b] 94.47 (1.61) | [c] 94.80 (3.49) | [d] **98.84** (0.79) |
| p values of paired t tests: [a] p = 0.0003, [b] p = 0.003, [c] p = 0.04, and [d] p = 0.001. | | | |

Results of Data Efficiency experiment. Figure 4 reports the effect of train set size on the dice scores of predicted masks on the independent test set. Compared to the results of the main experiment (N=586), we observed no drop in the overall dice scores by reducing the train set size suggesting that similar results could be achieved using much smaller cohorts. We also did not observe any consistent gain/loss by using both FLAIR and T1Gd as input to the Dense V-Net.

Table 2 – Comparison with other non-DL methods on the brain tumor test set. Values represent mean and standard deviation of scores.

| Method | Dice score $\mu(\sigma)$ | Sensitivity $\mu(\sigma)$ | Specificity $\mu(\sigma)$ | Average runtime per case |
|---|---|---|---|---|
| MONSTR | 91.34 (6.76) | [§d] 88.22 (7.44) | **98.91** (2.22) | 10-20 mins |
| SPM12-p | 93.36 (3.75) | [d¶] 93.39 (6.59) | 98.76 (1.05) | 2-3 mins |
| DeepBrain | **94.54** (1.09) | [§¶] **96.39** (2.34) | 98.48 (1.05) | **2-3 s** |
| p values of paired t tests: [§] p < 10⁻⁶, [¶] p = 0.022, and [d] p = 0.006. | | | | |

Performance on healthy cases. Table 3 presents the performance of our model on normal brains from the LBPA40[29] dataset. On average, DeepBrain achieved a dice of 96.2%, sensitivity of 96.6%, and specificity of 99.2% on this dataset. As Table 3 shows, our results are comparable to the state-of-the-art CNN approaches in the literature, however our dice score and sensitivity was on the lower end of these scores among the DL approaches. We believe this is expected given that unlike others we trained our model using brain-tumor patient data that includes unexpected brain appearance due to edema and necrosis. In comparison, the average turn-around time of our model for new test cases is drastically shorter than others.

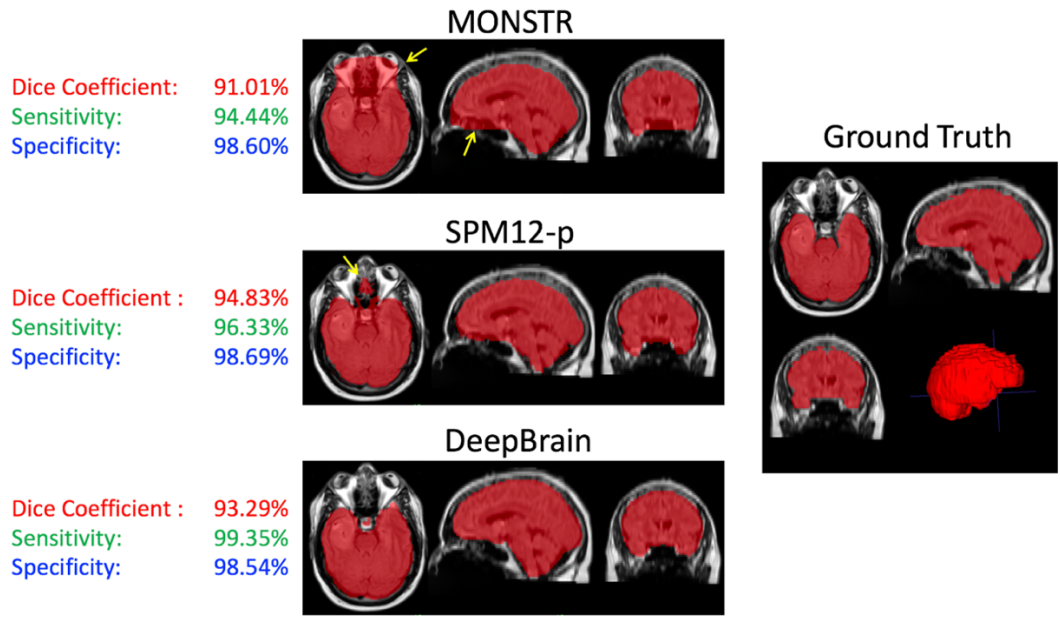

Figure 3 - Masks overlaid on the brain tumor MRIs; images on the left show the brain masks created using MONSTR, SPM12-p, and our DL model, DeepBrain in different anatomical views. The right image shows the ground truth manual segmentation. Our method performed very well and much better than the other methods in this application. The Dice coefficient, sensitivity, and specificity, calculated based on the ground truth for this case, are shown to the left of each image.

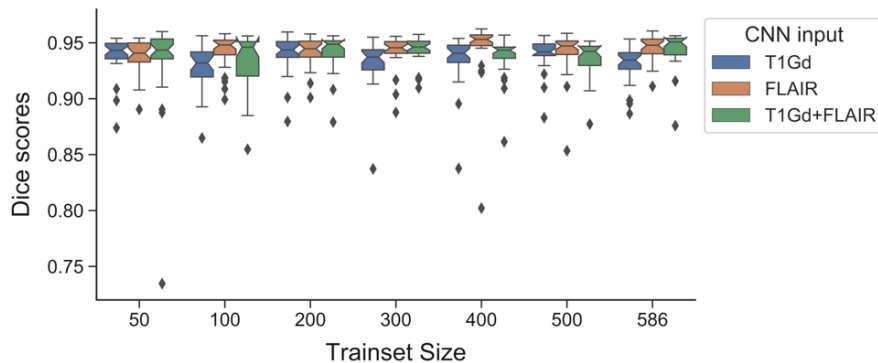

Figure 4 - The effect of training size on accuracy. Comparable dice scores were generated for the independent test set using models with various trainset sizes between 50 and 586. We didn't observe any consistent gain by using both FLAIR and T1Gd series as inputs.

Table 3 - Comparison with previous literature on manual segmentation of healthy brains in the LBPA40[29] dataset. Values for scores and run time in others' work are from literature.

| Method | Dice score $\mu(\sigma)$ | Sensitivity $\mu(\sigma)$ | Specificity $\mu(\sigma)$ | Average testing time (s) |
|---|---|---|---|---|
| CONSNet Lucena et al[57] | 97.35 (0.003) | 97.26 (0.007) | **99.54** (0.001) | 20 |
| Auto-U-net Salehi et al[40] | **97.73** (0.003) | **98.31** 0.006 | 99.48 (0.001) | 10.03 |
| U-Net Salehi et al[40] | 96.79 (0.004) | 97.22 (0.016) | 99.34 (0.002) | 4.57 |
| 3D CNN Kleesiek et al [41] | 96.96 (0.01) | 97.46 (0.01) | 99.41 (0.003) | 36.51 |
| Our approach | 96.17 (0.22) | 96.60 (0.08) | 99.22 (0.09) | **2** |

## Discussion

Despite the large body of existing literature on automatic skull striping methods on MRI, few have reported robustness in cases with a pathological brain. Among the non-learning based skull stripping approaches in the literature, the MONSTR algorithm[31] outperformed BEaST[23], SPECTRE[24], OPTIBET[30], and ROBEX[27] on a small cohort of 5 brain tumor cases with an average dice agreement of 96.95% with ground truth[31]. Unfortunately, our data did not support this level of performance for MONSTR and we achieved a moderate dice score of 91.34% and sensitivity of 88.22%. In comparison, the performance of SPM12-p was much better than MONSTR, particularly with respect to its much superior sensitivity (93.39%). This could be associated with the atlas-based nature of the MONSTR segmentation which results in inaccuracies when images deviate from healthy brain MRIs. Discrepancies could also be related to our use of T1Gd and FLAIR inputs to MONSTR, as the original results[31] were reported for T1 and T2. Given the large size of our cohort and the labor-intensive nature of manual segmentation, we needed an automatic method to create brain masks for training. We selected SPM12 due to its comparable performance with manual delineation of total intracranial volume in the presence of neurodegenerative pathology[45]. With minimal post-processing to compensate for the unexpected effects of the presence of tumor on images, our model achieved a dice score of 94.54% and sensitivity of 96.39% using only FLAIR images as input.

The closest work to ours is the modality-agnostic 3D convolutional neural network created by Thakur et al[59]. In this work, authors trained their network with pretreatment images of glioma patients using the common MR sequences in oncology including T1, T1Gd, T2, and FLAIR. Their model achieved an average dice coefficient of 97.8% on images from training institution and 95.6%, 91.6%, and 96.9% on datasets of other institutions. Our result is within the range of dice scores reported in their work for other institutes. Given the multi-centric nature of our brain tumor repository and the heterogeneity of its data, we believe the performance of our model is comparable to theirs. Moreover, one advantage of training on a heterogenous dataset with samples from many institutions is that it closely approximates the range of data found in clinical practice. Kleesiek at al[41] also reported an average dice of 95.2% and a sensitivity of 96.25% when testing on brain tumor data.

Our performance on healthy subjects was decidedly on the lower end of reported results for skull stripping deep learning models (Table 3). Salehi et al[40] compared the performance of a voxel-wise approach using three convolutional pathways for each anatomical plane, and a fully convolutional U-net[60] architecture and achieved dice coefficients of 97.7% and 96.8% on two publicly available datasets of normal brains. Kleesiek at al[41] trained a 3D input-agnostic fully convolutional network and compared its performance to 6 skull stripping methods (BET, BEaST, BSE, ROBEX, HWA, 3dSkullStrip) on publicly available datasets. Though the authors reported their performance in a merged public dataset, others[57] reported their performance on the LBPA40 dataset for a dice of 97.0% and sensitivity of 97.4%. Lucena et al[57] adopted a brain extraction method called CONSNet which consists of three parallel fully convolutional networks using the U-Net architecture and achieved a dice score of 97.3% and sensitivity of 97.2%. In this work, authors automatically generated silver standard labels for training using the STAPLE approach which combines 8 different segmentation approaches into a probabilistic consensus mask. Our approach did not yield the same level of accuracy on the LBPA40 dataset. We believe this is expected given that, unlike others, we trained our network using only brain-tumor patient data. Also, we trained our model using automatically-generated labels using only one method (SPM12-p).

We also conducted a data efficiency experiment in which training was repeated using progressively smaller cohorts. Our results demonstrated that, for the task of MRI skull stripping, a train set size of 50 MRIs might be sufficient to successfully train a convolutional neural network. Although larger datasets are always desirable, they are often unavailable in medical imaging. We thus suggest that rather than collecting large cohorts for training skull stripping CNN models, future efforts should focus on improving training labels and adopting an optimized learning approach.

Our work has a number of limitations. Firstly, we did not train a modality-agnostic model. Given the heterogeneity of data types that we observe across institutions, a modality-agnostic approach is necessary for ensuring utility across sites. Secondly, our training labels were generated using only one automatic method. Consensus methods have shown to be a more reliable alternative to any single automatic method in segmenting brain tissue [57]. In future work we aim to address both of these shortcomings.

## Conclusion

In this work we assessed the performance of a deep learning approach in extracting the brain on pretreatment MRI data of brain tumor patients acquired from over 20 institutions. We trained our network in a large cohort of patients using automatically-generated labels using the SPM12 software. Overall, our approach reached the highest accuracy with FLAIR images as input and our results on previously unseen brain tumor data was comparable to previous work in the literature. The data efficiency analysis showed that comparable levels of accuracy could have been achieved with 50 training samples. In conclusion, this study showed that whole brain extraction using deep learning approaches are more robust and accurate compared to alternative approaches and that comparable performance can be achieved with training on relatively smaller cohorts.

https://pdfs.semanticscholar.org/9e2d/7fea4957d79b81e1d035e9507b60d4c8c4a6.pdf.

Supplementary Table 1 - list of parameter settings in NiftyNet configuration file. Any parameter not specifically described here was left at the default value in the pipeline.

| Input: [T1Gd], [Flair], [Label] | | [Network] | |
|---|---|---|---|
| path_to_search | /path_to_data | name | dense_vnet |
| spatial_window_size | (144,144,144) | whitening | Ture |
| [System] | | batch_size | 6 |
| num_threads | 6 | window_sampling | resize |
| num_gpus | 1 | volume_padding_size | 0 |
| queue_length | 36 | window_sampling | resize |
| model_dir | /model_dir_path | optimiser | adam |
| data_split_file | /file_path | [Segmentation] | |
| [Training] | | Image | T1Gd + Flair |
| Sample per volume | 1 | label | label |
| Learning rate | 0.001 | Label normalization | False |
| Loss type | Dice | Num classes | 2 |
| max_iter | 500 | [Inference] | |
| save_every_n | 20 | border | (0, 0, 0) |
| starting_iter | 0 | spatial_window_size | (144, 144, 144) |
| | | save_seg_dir | /path_to_dir |